\documentclass[a4paper,11pt]{article}
\usepackage{pos}
\usepackage{subcaption}

\title{Hadronic susceptibilities for b to c transitions from two point correlation functions}

\author*[a]{Aurora Melis}
\author[a]{Francesco Sanfilippo}
\author[a]{Silvano Simula}

\affiliation[a]{Istituto Nazionale di Fisica Nucleare, Sezione di Roma Tre\\
Via della Vasca Navale 84, I-00146 Rome, Italy}

\emailAdd{aurora.melis@roma3.infn.it}

\abstract{We present a lattice determination of the hadronic susceptibilities that, thanks to unitarity and analyticity, constrain the form factors entering the semileptonic $b\rightarrow c$ transitions. We evaluate the transverse and longitudinal  susceptibilities of the vector and axial polarization functions at zero momentum transfer from the moments of appropriate two-point correlation functions. The latter are obtained on the lattice employing gauge ensembles of the Extended Twisted Mass Collaboration (ETMC) with $N_f$=2+1+1 flavors of Wilson-clover twisted-mass quarks with masses of all the dynamical quark flavors tuned close to their physical values. The simulations are carried out at four values of the lattice spacing, $a \simeq (0.057,0.068, 0.080, 0.091)$~fm, with spatial lattice sizes up to $L \simeq 7.6$~fm. Heavy-quark masses up to $\approx 3.5$ times the physical charm mass are employed, allowing for a smooth extrapolation to the physical b-quark mass.}

\FullConference{The 40th International Symposium on Lattice Field Theory (Lattice 2023)\\
July 31st - August 4th, 2023\\
Fermi National Accelerator Laboratory\\}

\begin{document}

\maketitle
\section{Introduction}
The hadronic form factors (FFs) entering the exclusive semileptonic $B$-meson decays $B\rightarrow D^{(*)} \ell \nu_\ell$ are crucial ingredients for the extraction of the CKM matrix element $|V_{cb}|$ from experimental data and for a pure theoretical prediction of the lepton flavor universality ratios $R_{D^{(*)}} = BR(B\rightarrow D^{(*)}\tau \nu_\tau) / BR(B\rightarrow D^{(*)} e(\mu) \nu_{e(\mu)})$. 

The required hadronic FFs are non perturbative quantities that must be computed using lattice QCD simulations. Presently, the $B\rightarrow D^{(*)} \ell \nu_\ell$ FFs have been determined on the lattice with a few-percent accuracy for values of the 4-momentum transfer $q^2$ close to $q^2 \sim q^2_{max} \equiv (m_B - m_{D{(*)}})^2$, whereas experimental measurements are more precise in the region $q^2 \sim 0$. In order to predict the $q^2$-dependence of the FFs in the full semileptonic region, $0\lesssim q^2 \leq q^2_{max}$, truncated $z$-expansions are often employed mixing also theoretical and experimental data. All that, however, can bias the determinations of $|V_{cb}|$ and $R_{D^{(*)}}$. 

Recently, the dispersive matrix (DM) approach \cite{DiCarlo:2021dzg, Martinelli:2021onb, Martinelli:2021myh, Martinelli:2023fwm} has been proposed as a tool to predict the full $q^2$-dependence of the FFs in a model-independent way based only on lattice QCD results, unitarity and analyticity. 
The starting point of the method is a dispersive bound that, for a generic form factor $f$, reads
\begin{eqnarray}
    \chi(q^2)\geq \frac{1}{\pi}\int^\infty_{q_{thr}^2} dt \frac{W(t)|f(t)|^2}{(t-q^2)^3},
\end{eqnarray}
where $W(t)$ is a known phase space factor, $q_{thr}^2$ the relevant annihilation threshold and $\chi(q^2)$ is the hadronic \emph{susceptibility}. The latter are defined as moments of suitable two-point functions $\Pi^{JJ}_{\mu\nu}$ that for a flavor-changing current $J_\mu = V_\mu, A_\mu$ $(V_\mu=\overline{c}\gamma_\mu b$ and $A_\mu=\overline{c}\gamma_5\gamma_\mu b)$ splits into a transverse ($T$) and longitudinal ($L$) polarization functions:
\begin{eqnarray}
    \label{eq:Pimunu}
    &\Pi^{JJ}_{\mu\nu}(q)
    \equiv i\int d^4x e^{iq\cdot x} \langle0| T \{J_\mu (x) J^\dagger_\nu(0)\}|0\rangle = (q_\mu q_\nu-g_{\mu\nu}q^2)\Pi^{JJ}_T(q^2) + q_\mu q_\nu \Pi^{JJ}_L(q^2) ~ , ~ \\[2mm]
    \label{eq:Chi_def}
    &\chi^{J}_T(q^2) \equiv \frac{1}{2} \dfrac{\partial^2[q^2\Pi^{JJ}_T(q^2)]}{(\partial q^2)^2} ~ , ~ \qquad \chi^{J}_L(q^2) \equiv \dfrac{\partial [q^2\Pi^{JJ}_L(q^2)]}{\partial q^2}.
\end{eqnarray}
Performing a Wick rotation from Minkowskian to Euclidean coordinates, one can express the polarization functions in terms of the 2-point Euclidean correlation functions $C(\tau)$ as \cite{Martinelli:2021frl}
\begin{eqnarray}
\label{eq:PiTL_Ct}
    Q^2\Pi^{JJ}_{T}(Q^2) = - \int^{+\infty}_{-\infty} d\tau e^{-iQ\cdot \tau} C^{J_i J_i}(\tau) ~ , ~  \quad Q^2\Pi^{JJ}_{L}(Q^2) = - \int^{+\infty}_{-\infty} d\tau e^{-iQ\cdot \tau} C^{J_0 J_0}(\tau) ~ , ~
\end{eqnarray}
where $Q^2 = - q^2$ and the Euclidean four-momentum $Q$ is chosen in the temporal direction $(Q, \vec{0})$.
The first nonperturbative determination of the hadronic susceptibilities for $b\rightarrow c$ transitions has been carried out in Ref.~\cite{Martinelli:2021frl} using $N_f=2+1+1$ gauge configurations produced by the Extended Twisted Mass Collaboration (ETMC).
To improve that analysis, in our calculations we make use of the latest $N_f=2+1+1$ ETMC gauge configurations \cite{ExtendedTwistedMass:2021gbo}, which differ from those used in Ref.\,\cite{Martinelli:2021frl} (i) for the use of Wilson-clover twisted-mass quarks, (ii) for considering four values of the lattice spacing $(a\simeq 0.057,0.068,0.080,0.091)$~fm, and (iii) for having all dynamical quark masses at their physical point on most ensembles.

As for the 2-point correlation functions, we use two regularizations of the twisted mass formulation, namely with equal $(r,r)$ or opposite $(r,\text{-}r)$ values of the Wilson parameter $r$, which differ by ${\cal{O}}(a^2)$ effects. To reduce the discretization effects, we subtract the leading order (LO) cut-off effects evaluated in the free theory. Finally in order to extrapolate to the physical $b$-quark we simulate a series of $\lambda$-spaced heavy quark masses $m_h(n) = \lambda^n m_c$ with $n = 1 - 8$ and $\lambda \approx 1.16$,  leading to $m_h \lesssim 3.5 m_c \approx 0.75 m_b$.
For each ensemble we analyze the first and second derivatives appearing in Eq.\,(\ref{eq:Chi_def}) for the transverse and longitudinal polarization functions of both the vector and axial currents. At zero momentum transfer one has \cite{Martinelli:2021frl}
\begin{eqnarray}
\label{eq:Chi20}
   & m^2_h \chi^{V}_T (a^2;m_h,m_c) = \frac{m^2_h}{12}\int^{\infty}_{a}d\tau\,\tau^4 C^{V_i V_i}(\tau) ~ , ~ 
                  \chi^V_L(a^2;m_h,m_c) = \frac{(m_h-m_c)^2}{12}\int^{\infty}_{a}d\tau\,\tau^4 C^{SS}(\tau) ~ , ~ \\[2mm]    
   &m^2_h \chi^{A}_T (a^2;m_h,m_c) = \frac{m^2_h}{12}\int^{\infty}_{a}d\tau\,\tau^4 C^{A_i A_i}(\tau) ~ , ~ 
                 \chi^A_L(a^2;m_h,m_c) = \frac{(m_h+m_c)^2}{12}\int^{\infty}_{a}d\tau\,\tau^4 C^{PP}(\tau) ~ , ~ \nonumber
\end{eqnarray}
where we have exploited the Ward Identities (WIs) to express the longitudinal susceptibilities as the fourth moments of the scalar ($S$) and pseudoscalar ($P$) correlation functions. The (dimensionless) susceptibilities in Eq.~(\ref{eq:Chi20}) has to be extrapolated to the continuum limit and to the physical $b$-quark mass, as it will be described in the next sections.

\section{Physical b-quark point}
In order to extrapolate the susceptibilities to the physical $b$-quark point we make use of the simulated mass of the $B_c$ meson, obtained from the pseudoscalar correlation function used to determine $\chi^A_L$. We study also the $B^*_c$ and $B_s$ meson masses as a consistency check. The ground-state mass of  $P$ and $V$ mesons, $M_{P(V)}$, are extracted from a constant fit to the plateau of the effective mass $M^{\rm eff}_{P(V)}$ at large time distances from the source.

We take the starting value $m_c$ obtained in Ref.\,\cite{ExtendedTwistedMass:2021gbo} for the charm-quark mass. For this value we obtain $M_P(m_c, m_c) = 2981(18)$~MeV, $M_V(m_c, m_c) = 3095(18)$~MeV and $M_P(m_c, m_s) = 1964(12)$~MeV, which are compatible with the more precise experimental values \cite{ParticleDataGroup:2022pth} $M^{\rm exp}_{\eta_c} = 2983.9(4)$~MeV, $M^{\rm exp}_{J/\Psi} = 3096.900(6)$~MeV and $M^{\rm exp}_{D_s} = 1968.35(7)$~MeV, respectively.

\begin{figure}[htb!]
    \centering
    \includegraphics[width=0.80\textwidth]{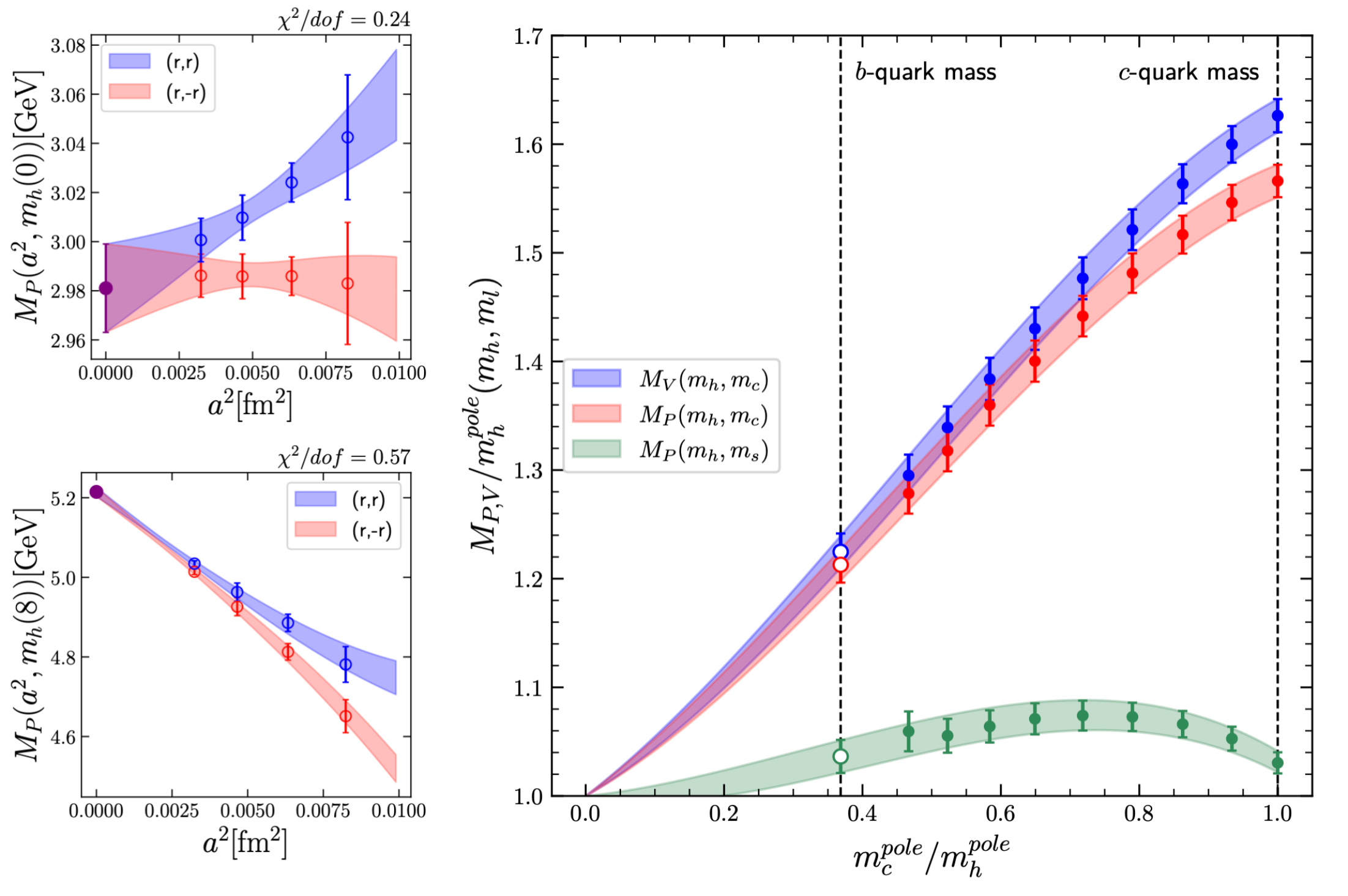}
    \vspace{-5mm}
    \caption{ (Left panels) Continuum limit extrapolation of the meson mass $M_P(m_h,m^{\rm phys}_c)$ for the lightest and heaviest $m_h(n)$ considered. The blue and red points correspond to the $(r,r)$ and $(r,\text{-}r)$ regularizations. The colored bands represent the results of the fit (\ref{eq:a2fit_MVP}). (Right panel) Meson masses versus the heavy-quark mass ratio $m_c/m_h$ in the pole mass scheme. The colored bands represent the results obtained with the fitting function of Eq.(\ref{eq:Mfitmh}). The white dots are the fit values at the physical b-quark mass, indicated by vertical dotted line.}
    \label{fig:Fig1}
\end{figure}

We extrapolate the masses to the the continuum limit adopting a combined fit of the results corresponding to the two regularizations $(r,r)$ and $(r,\text{-}r)$, imposing the same continuum limit. We study two linear combinations of the original data sets, namely the mean and the difference. We observe that discretization effects on the average of the two regularizations nicely scale with $a^2$. Thus, the two regularizations are fitted using the following polynomial Ansatz
\begin{equation}
    \label{eq:a2fit_MVP}
    M^{(r, \pm r)}(a^2; m_h, m_c) =  M(m_h, m_c) \left( 1 + A^{(r, \pm r)}_1 a^2 \pm A_2 a^4  \pm A_3 a^6 \right) ~ , ~
\end{equation}
that includes a total of 5 free parameters for each heavy-quark mass $m_h$: $M(m_h,m_c), A^{(r,r)}_{1},A^{(r,\text{-}r)}_{1}$ and $A_{2,3}$. The quality of the fits is shown in Fig.~\ref{fig:Fig1} (right panel), where we show also separately the fits for the two regularizations in the case of the lightest and heaviest quark mass considered. 
To extrapolate to the physical $b$-quark mass we exploit the fact that the mass of a heavy meson must equal the pole mass of its heavy quark in the static limit, namely 
 $\lim_{m_h\rightarrow\infty} M_{P(V)}(m_h,m_{c(s)}) / m^{pole}_h = 1$.
For both P and V channels we employ a simple ansatz including 3 free parameters $B_{1,2}$ and $B^s_2$:
\begin{eqnarray}
    \label{eq:Mfitmh}
    \frac{M(m_h,m_c)}{m^{pole}_h} = 1 + B_1 \frac{m^{pole}_c}{m^{pole}_h} + \left(B_2 + B^s_2 \frac{\alpha_s(m^{pole}_h)}{\pi}\right)\left(\frac{m^{pole}_c}{m^{pole}_h}\right)^2.
\end{eqnarray}
The quality of the fit is shown in Fig.~\ref{fig:Fig1} (left panel) for the cases $M = M_{P(V)}(m_h,m_c)$ and $M_P(m_h,m_s)$. 
We tune the value of $\lambda$ to reproduce the experimental $B_c$ meson mass, $M^{\rm exp}_{B_c} = 6274.47(33)$~MeV \cite{ParticleDataGroup:2022pth}, with $n =10$ steps in $\lambda$. In this way we obtain
\begin{eqnarray}
\label{eq:lam_mb}
   \lambda=1.1652(10), \qquad m^{\overline{\rm MS}}_b/m^{\overline{\rm MS}}_c \equiv \lambda^{10} = 4.612(40).
\end{eqnarray}
The result for the $b/c$ quark mass ratio is  compatible within one standard deviation with the result of Ref.~\cite{Hatton:2021syc}.
Moreover, we obtain the following predictions for others $b$-quark mesons: $M_{B_s}=5364(48)$~MeV and $M_{B^*_c}=6327.3(47)$~GeV that nicely compare with $M^{\rm exp}_{B_s} = 5366.92(10)$~MeV \cite{ParticleDataGroup:2022pth} and $M_{B^*_c} = 6331(7)$~MeV from Ref.\,\cite{Mathur:2018epb}.

\section{Susceptibilities on the lattice}

\subsection{Perturbative subtraction and continuum limit }

\begin{figure}[htb!]
    \centering
    \includegraphics[width=\textwidth]{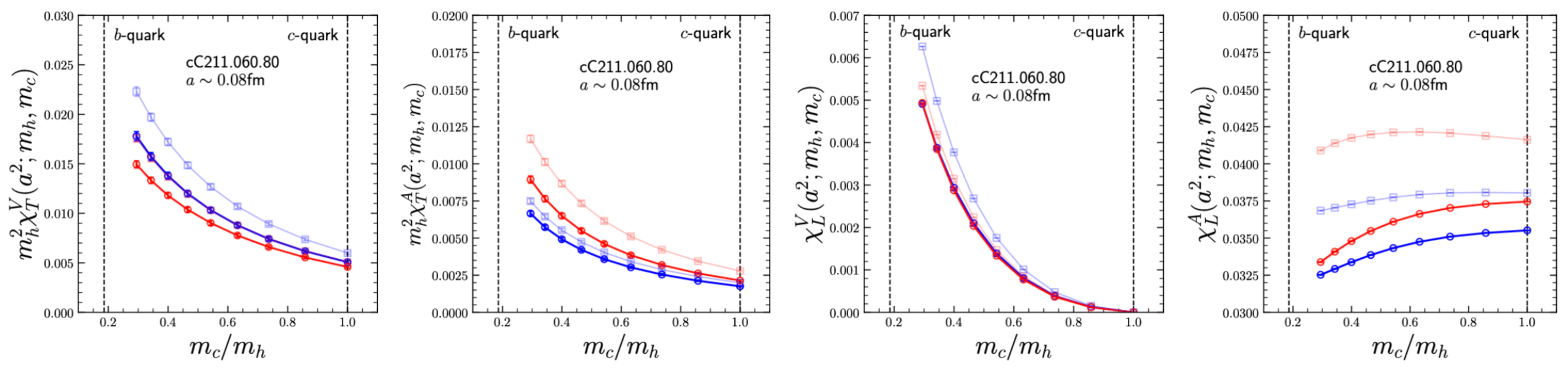}
    \vspace{-7mm}
     \caption{Heavy-quark mass dependence of the susceptibilities in different channels for a given ensemble. The shaded and full markers correspond respectively to the susceptibilities  before and after the perturbative subtraction of the LO discretization effects for the $(r, r)$ (blue) and $(r, \text{-}r)$ (red) regularizations.}
    \label{fig:Fig2}
\end{figure}

Following the procedure of Refs.~\cite{DiCarlo:2021dzg, Martinelli:2021frl} to improve the continuum limit, we subtract from the non-perturbative simulated susceptibilities in Eq.~(\ref{eq:Chi20}) the difference between the susceptibilities in the free theory (FT), computed at each simulated lattice spacing, and the continuum value from perturbation theory (PT) at LO from Ref.~\cite{Boyd:1997kz}, namely
\begin{eqnarray}
\label{eq:Chisub}
    \chi_{T,L}(a^2;m_h,m_c) &\rightarrow& \chi_{T,L}(a^2;m_h,m_c) - \left(\chi^{\rm FT}_{T,L}(a^2;m_h,m_c)-\chi^{\rm PT}_{T,L} (m_c/m_h)\right) ~ , ~
\end{eqnarray}
where to streamline the notation we omit the dependence on the particular current.
By construction, the bracket in Eq.~\ref{eq:Chisub} piece contains the discretization effects present in the free theory for our lattice setup. The procedure is beneficial in decreasing the discretization effects as shown in Fig.~\ref{fig:Fig2}, where the difference between the two regularisations $(r, \pm r)$ is clearly reduced for each channel. 
\begin{figure}[htb!]
    \centering
    \includegraphics[width=\textwidth]{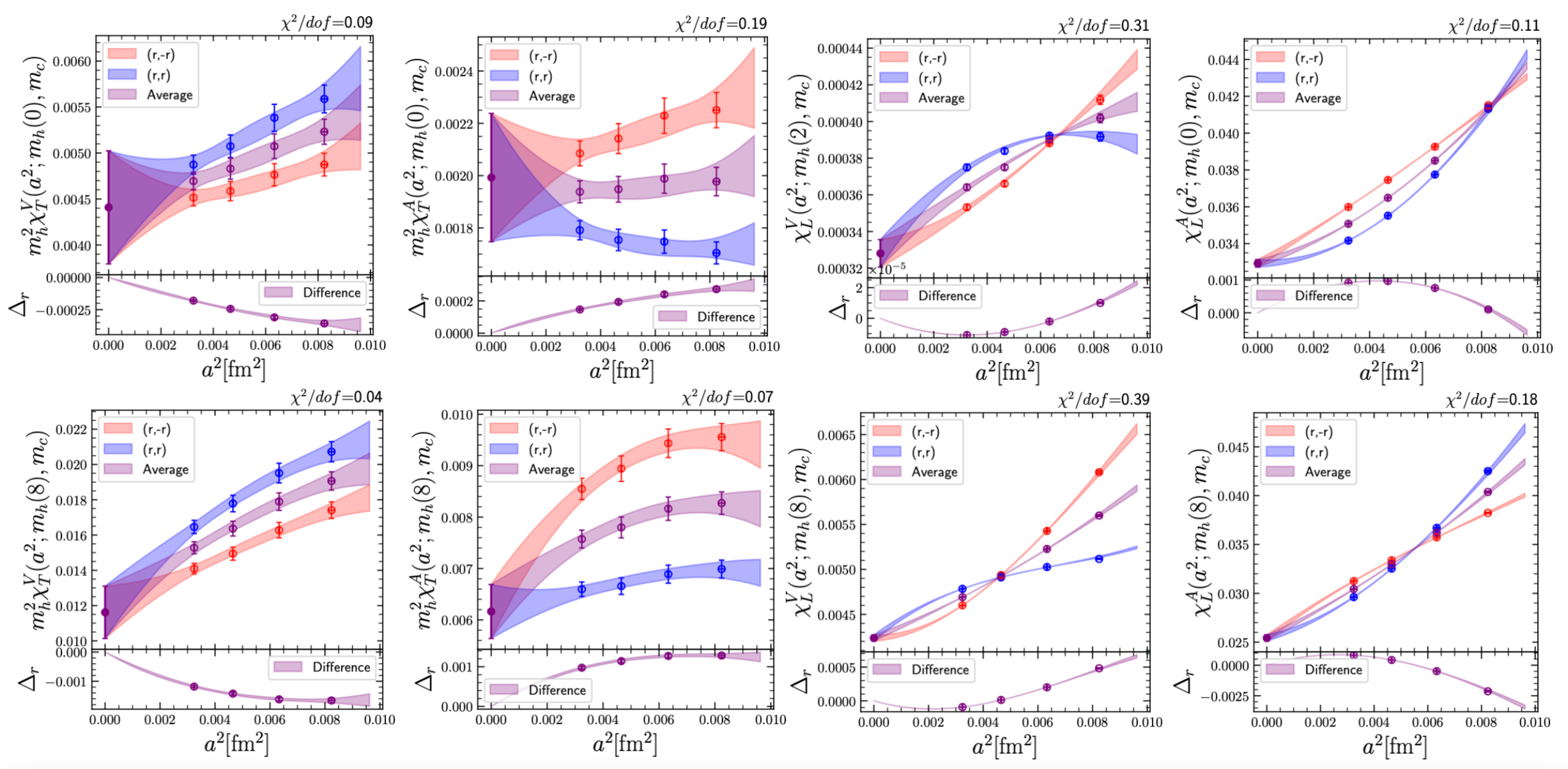}
    \vspace{-7mm}
    \caption{Continuum limit extrapolations of the susceptibilities in the different channels for the lightest $m_h(0) = m_c$ and heaviest mass considered $m_h(8) \simeq 3.5 m_c$ (right panels). The blue (red) points correspond to the $(r, r)$ and $(r, \text{-}r)$ regularizations, while the violet points display their average and difference (bottom panels). The darker colored bands represent the results obtained with the fitting function (\ref{eq:a2fit_Chi}).}
    \label{fig:Fig3}
\end{figure}

We perform the continuum extrapolation adopting an Ansatz analogous to Eq.\,(\ref{eq:a2fit_MVP}), i.e.
\begin{equation}
    \label{eq:a2fit_Chi}
    \left[ \chi^J_{T(L)} \right]^{(r, \pm r)} (a^2; m_h, m_c) = \chi^J_{T(L)} (m_h, m_c)  \left( 1 + C^{ (r, \pm r)}_1 a^2  + C^{(r, \pm r)}_2 a^4 \pm C_3 a^6 \right) ~ , ~
 \end{equation}
which contains again a total of five free parameters for each heavy-quark mass $m_h$.
The quality of the fits is shown in Fig.~\ref{fig:Fig3} for the lightest and  heaviest quark mass considered. We present the two regularizations together with their average and difference $\Delta_r$. In most of the cases the mean displays a simpler dependence in $a^2$, while the main $\mathcal{O}(a^4)$ and $\mathcal{O}(a^6)$ effects are present in the difference.

\subsection{Extrapolation at the b-quark point}
\begin{figure}[htb!]
    \centering    
    \includegraphics[width=\textwidth]{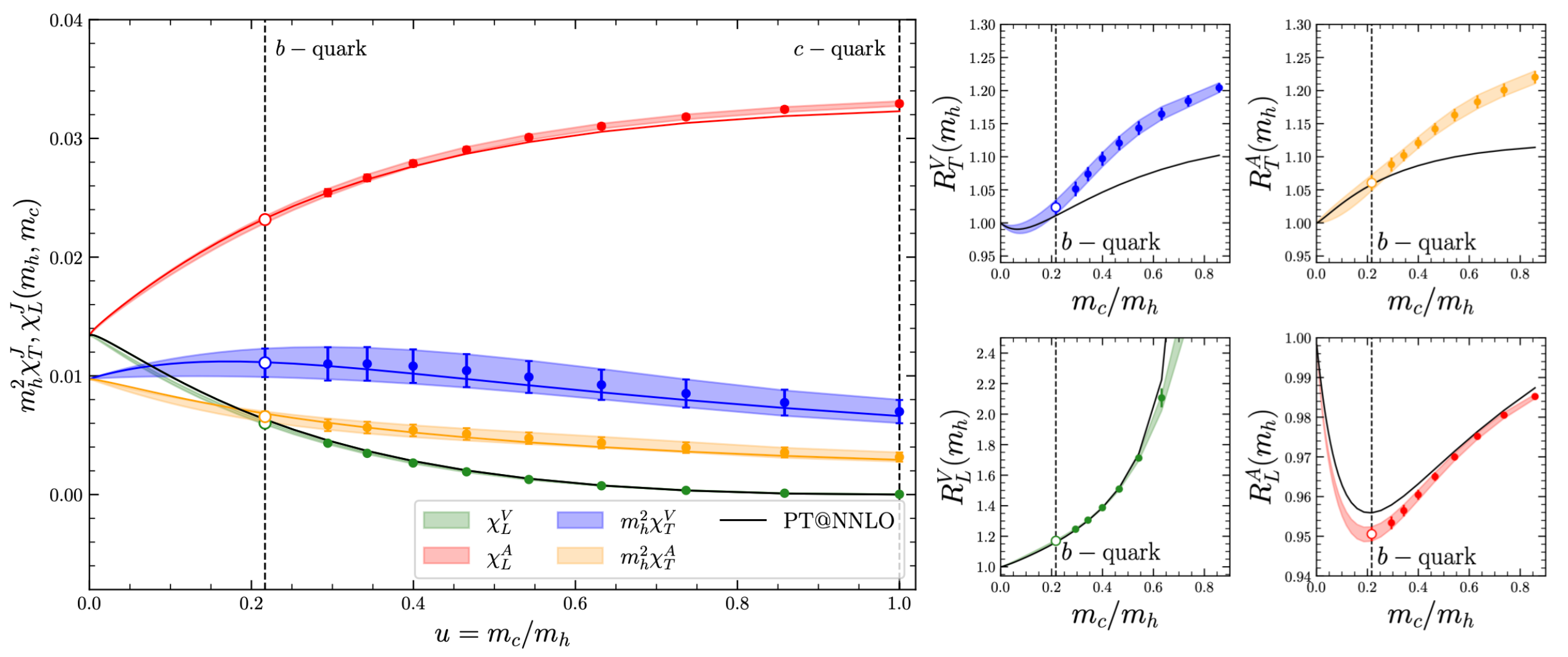}
    \vspace{-7mm}
    \caption{The susceptibilities $m^2_h\chi^J_{T}$ and $\chi^J_{L}$ (left panel) and the ratios $R^{J}_{T,L}$ (right panels) at $q^2=0$ in the different channels versus the inverse heavy-quark mass ratio $m_c/m_h$. The colored bands represent the results obtained with the fitting function in Eq.~(\ref{eq:bfit1}). The white dots are the fit values at the physical $b$-quark point, indicated by vertical dotted lines. The black lines are the PT expressions at NNLO taken from Ref.~\cite{Grigo:2012ji}.}
    \label{fig:Fig4}
\end{figure}

The susceptibilities are extrapolated to the physical $b$-quark mass following two approaches: ($1^{st}$) a direct fit of the susceptibilities and ($2^{nd}$) the \emph{ETMC ratio method}. 
The first approach relies on the fact that the Operator Product Expansion (OPE) analysis of these quantities (see Ref.~\cite{Boyd:1997kz}) predicts the LO non-perturbative corrections to be suppressed by $1/m^4_h$ and therefore absent in the static limit. This allows to fit the four susceptibilities imposing the same static limit of the PT:
\begin{eqnarray}
    &&\lim_{m_h\rightarrow \infty} m^2_h\,\chi^J_T(m_h,m_c) = 3/32\pi^2 + 0.0108309\,\alpha_s + 0.0120914\,\alpha^2_s + \mathcal{O}(\alpha^3_s),\nonumber\\
    &&\lim_{m_h\rightarrow \infty} \chi^J_L(m_h,m_c) = 1/8\pi^2 + 0.00951385\, \alpha_s + 0.0079923\,\alpha^2_s + \mathcal{O}(\alpha^3_s),
\end{eqnarray}
where the $\mathcal{O}(\alpha^2_s)$ coefficients are known from Ref.~\cite{Grigo:2012ji} in the $\overline{\rm MS}$-scheme at the scale $\mu = m_h$.
We use the following fitting function that contains a total of 6 free parameters, $D^{(s)}_{1,2,3}$, i.e.
\begin{eqnarray}
    \label{eq:bfit1}
    \frac{\chi^J_{T(L)}(m_h,m_c)}{\chi^{PT}_{T(L)}(0)} &=& 1 + \sum^3_{k=1} \left[ D_k + D^s_k \frac{\alpha_s(m_h)}{\pi}\right]\left(\frac{m_c}{m_h}\right)^k .
\end{eqnarray}
The quality of the fits is displayed in Fig.~\ref{fig:Fig4} together with the expressions from PT at NNLO \cite{Grigo:2012ji}.

The second approach adopts the \emph{ETMC ratio method} \cite{ETM:2009sed}. To this end we build ratios of susceptibilities at subsequent heavy-quark masses 
\begin{eqnarray}
    R^J_{T}(m_h(n)) \equiv
    \frac{m^2_h(n)}{m^2_h(n-1)} \frac{\chi^J_T(m_h(n),m_c)}{\chi^J_T(m_h(n-1),m_c)} ,\quad
    R^J_{L}(m_h(n)) \equiv
    \frac{\chi^J_L(m_h(n),m_c)}{\chi^J_L(m_h(n-1),m_c)}.  
\end{eqnarray}
In this way no assumption on the susceptibilities in the static limit is required, since $\lim_{m_h\rightarrow\infty} R^{J}_{T(L)}(m_h) = 1$.
In this analysis we fit directly each ratio $R^J_{T(L)}$ using the fitting function similar to Eq.~(\ref{eq:bfit1}).
The susceptibilities at the physical $b$-quark point can be obtained from the chain of products
\begin{eqnarray}
   \{ m^2_h \chi^J_T, \chi^J_L \}(m_b,m_c) = \{ m^2_h \chi^J_T, \chi^J_L \}(m_h(n_{\rm trig}-1))\prod^{n_b}_{n=n_{\rm trig}} R^J_{T,L} (m_h(n)) ~ , ~
\end{eqnarray}
where $n_{\rm trig} = 2$ for $\chi^{V}_L$ (since by charge conservation $\chi^{V}_L(m_c, m_c) = 0$) and $n_{\rm trig }=1 $ for all the other channels.
We quote our preliminary results for the susceptibilities at $Q^2 = 0$:
\begin{equation}
    \nonumber
    \begin{tabular}{c c}
    $1^{st}$ analysis           & $2^{nd}$ analysis           \\[5pt]
    \hline \\[-10pt]
    $\chi^V_T(m_b,m_c) = 5.6(6)\times 10^{-4}$ GeV$^{-2}$ &  $\chi^V_T(m_b,m_c) = 5.8(8)\times 10^{-4}$ GeV$^{-2}$ \\[5pt]
    $\chi^V_L(m_b,m_c) = 6.0(1) \times 10^{-3} $ &  $\chi^V_L(m_b,m_c) = 6.0(2) \times 10^{-3}  $\\[5pt]
    $\chi^A_T(m_b,m_c) = 3.3(2)\times 10^{-4}$ GeV$^{-2}$ &  $\chi^A_T(m_b,m_c) = 3.3(3)\times 10^{-4}$ GeV$^{-2}$\\[5pt]
    $\chi^A_L(m_b,m_c) = 2.32(3)\times 10^{-2} $ &  $\chi^A_L(m_b,m_c) = 2.30(4)\times 10^{-2} $
    \end{tabular}
\end{equation}
The two analysis provide compatible results. Our present results suggest that non-perturbative condensate terms in the relevant OPE are small.

\end{document}